\documentclass[runningheads]{llncs}

\usepackage{amsmath,amssymb,amsfonts}
\usepackage{algorithmic}
\usepackage{graphicx}
\usepackage{textcomp}
\usepackage[dvipsnames]{xcolor}
\usepackage{color,soul}
\usepackage[numbers]{natbib} 

\usepackage[hyphens]{url}



\usepackage{multirow, multicol}
\usepackage{makecell}

\let\oldcite\cite

\renewcommand{\cite}[1]{%
  \ifx\relax#1\relax%
    \textcolor{red}{$\langle$empty citation$\rangle$}%
  \else%
    \if\relax\detokenize{#1}\relax%
      \textcolor{red}{$\langle$empty citation$\rangle$}%
    \else%
      \oldcite{#1}%
    \fi%
  \fi%
}

\usepackage[hidelinks]{hyperref}
\usepackage{cleveref}
\usepackage{xspace}
\usepackage{balance}
\usepackage[font=footnotesize]{subcaption}
\usepackage{booktabs}
\usepackage{siunitx}

\def\BibTeX{{\rm B\kern-.05em{\sc i\kern-.025em b}\kern-.08em
    T\kern-.1667em\lower.7ex\hbox{E}\kern-.125emX}}

\def\eg{\emph{e.g.},\xspace} 

\def\ie{\emph{i.e.},\xspace}

\newboolean{showcomments}
\setboolean{showcomments}{false}

\ifthenelse{\boolean{showcomments}}
  {\newcommand{\nb}[2]{
    \fbox{\bfseries\sffamily\scriptsize#1}
    {\sf\small$\blacktriangleright$\textit{#2}$\blacktriangleleft$}
   }
    \newcommand{\rev}[1]{{\leavevmode\color{blue}#1}}
    \newenvironment{revfloatenv}
      {\par\noindent\color{blue}\rule{\linewidth}{1pt}\par\color{black}\vspace{.5ex}}
      {\vspace{-.5ex}\par\noindent\color{blue}\rule{\linewidth}{1pt}\par\color{black}}
  }
  {\newcommand{\nb}[2]{}
    \newcommand{\rev}[1]{{\leavevmode#1}}
}

\newcommand\ADP[1]{\textcolor{RawSienna}{\nb{ANDRES}{#1}}}

\begin{document}


\title{A Surrogate-based Approach for Fast Multi-objective Architectural Refactoring Optimization}
\titlerunning{Surrogate-based Fast Multi-objective Architecture Refactoring Optimization}

\author{J. Andres Diaz-Pace\inst{1} \and Daniele {Di Pompeo}\inst{2} \and Antonela Tommasel\inst{1,3}}

\institute{ISISTAN, CONICET-UNCPBA, Tandil, Argentina \and
SPENCER Lab, University of L'Aquila, L'Aquila, Italy \and
Johannes Kepler University Linz, Austria}

\maketitle

\begin{abstract}

Software model optimization is a process that generates architecture alternatives aimed at improving quantifiable non-functional properties of software systems, such as performance and reliability. 
Multi-objective evolutionary algorithms are commonly used to explore the search space and help designers identify trade-offs among competing non-functional properties (e.g., through a Pareto front). 
However, such algorithms face efficiency challenges in complex software models and large design spaces, since evaluating the fitness (i.e., the quality) of each architecture requires analysis tools that become computationally expensive when repeatedly invoked during the search process. 
In this paper, we explore the construction of surrogate models based on regression techniques to approximate the outputs of these analysis tools at significantly lower computational cost, while maintaining reasonable output accuracy. 
Our experimental results suggest that surrogate models provide savings of up to $30\%$ in computational time and maintain the Pareto front quality provided by evolutionary algorithms. Also, we observed some differences in the architectural models produced by our approach. Overall, surrogate models constitute a promising approach for scaling multi-objective architecture optimization to larger spaces and complex architectural models.

\end{abstract}


\section{Introduction}\label{sec:intro}\vspace{-0.3cm}

A software architecture is a high-level abstraction of a system that defines its internal structure in terms of components, connectors, properties and main functions. 
An architecture plays a crucial role in determining the blueprint of the software system and the non-functional properties that it should satisfy, such as maintainability, scalability, or performance, among others~\cite{DBLP:journals/infsof/NiDYMYHX21,Meedeniya-Buhnova-Aleti-Grunske-2010}.
Software architectures have been exploited to analyze non-functional properties of software systems~\cite{DBLP:conf/icsa/ArcelliCDP18, DBLP:conf/icsa/BuschFK19, Rago-Vidal-Diaz-Pace-Frank-van-Hoorn-2017}.
These efforts normally rely on predefined analysis models (or solvers), such as Layered Queuing Networks (LQNs) for performance~\cite{franks2008enhanced} or rule-based detection for performance antipatterns~\cite{Cortellessa-Di-Pompeo-2021}. 
Beyond analysis, architectural refactoring has emerged as a complementary activity aimed at systematically modifying the architecture structure (\eg by redistributing components, introducing replicas, or restructuring communication paths) to improve its quality attributes~\cite{DBLP:journals/software/Zimmermann15,DBLP:conf/icsa/ZhaoMB25,DBLP:journals/infsof/NiDYMYHX21}.
For example, \citet{DBLP:conf/icsa/ZhaoMB25} analyzed the energy efficiency of microservices of different architectural decompositions, 
and \citet{DBLP:journals/infsof/NiDYMYHX21} improved the performance of an architecture using rule-based refactoring.

When competing properties must be optimized together, multi-objective techniques{|}such as genetic algorithms~\cite{Meedeniya-Buhnova-Aleti-Grunske-2010,Martens-Koziolek-Becker-Reussner-2010}{|}are well suited to this task, particularly when the properties can be expressed as numerical quantities. 
However, non-functional properties are often challenging to optimize due to the complexity of the architectural models, the computational cost of running the analysis solvers, and the large search space~\cite{DBLP:conf/icsa/ArcelliCDP18,DBLP:journals/infsof/NiDYMYHX21}.
A common strategy to mitigate computational cost is to impose time budgets on the optimization process~\cite{diaz-paceRoleSearchBudgets2025,Arcuri-Fraser-2013}. 
Although effective for execution time, such constraints may hinder the exploration of the design space, affecting the quality of architectural alternatives. 

In this work, we address the high computational cost of evaluating the optimization objectives via solvers, 
which hampers multi-objective architectural optimization in practice.
We propose a novel \textit{approach based on surrogate models to estimate non-functional properties of architectures}, approximating the behavior of the solvers with cheaper Machine Learning models.
In particular, our approach leverages regression techniques to build surrogate models based on features extracted from the architectures handled by the optimization engine.

We evaluate our approach on \textit{CoCoME}~\cite{Herold-Klus-Welsch-Deiters-Rausch-Reussner-Krogmann-Koziolek-Mirandola-Hummel}, which is a well-known case study in the literature.
Furthermore, we compare the surrogate results with those achieved by a state-of-the-art search-based  technique~\cite{DBLP:conf/icsa/CortellessaP024}. 
Through this investigation, we aim to answer the following research questions: 

\vspace{0.15cm}
\noindent\textbullet~\textbf{RQ1:} Can surrogate models reduce the overall execution time of multi-objective architectural refactoring optimization while maintaining solution quality?

\noindent\textbullet~\textbf{RQ2:} Are there differences in the structural characteristics of the architecture candidates generated by a surrogate-assisted approach and those generated by a standard search-based optimization engine?

\vspace{0.15cm}

Our results show that surrogate models can significantly reduce the evaluation time of non-functional properties ($\approx 30\%$) during the optimization process, leading to faster convergence. 
A trade-off we observed here is less variability of architectural models in the exploration of the design space, as outlined by quality indicators employed to evaluate quality of Pareto fronts. With regard to the architectural models, we noticed that surrogate models lead to different models (in the Pareto front) than those generated by a standard evolutionary search.

\vspace{-0.5cm}\section{Related Work}\label{sec:related}\vspace{-0.3cm}

Although multi-objective optimization techniques, such as evolutionary algorithms, can yield good-quality and diverse solutions in different engineering domains, a main drawback is their execution time until reaching convergence.

Along this line, the definition of stopping criteria for the optimization process is a common strategy explored in the literature. 
\citet{Arcuri-Fraser-2011,Arcuri-Fraser-2013} showed that fixed time budgets influence search-based test generation, while \citet{Luong-Nguyen-Gupta-Rana-Venkatesh-2021} proposed a cost-aware Bayesian optimization method that explicitly incorporates the evaluation cost of candidate solutions. 
In the context of software architecture optimization, \citet{diaz-paceRoleSearchBudgets2025} examined how time budgets affect multi-objective refactoring search. 
Their findings indicate that, although budget-aware search reduces overall evaluation time, it may also lead to suboptimal architectural trade-offs depending on the search strategy, highlighting the sensitivity of architecture-level optimization to budget constraints. 

Surrogate models offer a complementary strategy by learning fast approximations of expensive evaluation functions. 
Their effectiveness has been demonstrated in surrogate-assisted evolutionary algorithms \cite{DBLP:journals/ec/HildebrandtB15}, where predictions replace a portion of fitness evaluations to accelerate convergence on expensive multi-objective problems.
Recent research has examined surrogate-based optimization specifically for system architecture problems, which are characterized by hierarchical, mixed-discrete, and multi-objective design spaces. 
Prior work has proposed Bayesian optimization with specialized structure~\cite{archopt_strategies}, as well as methods that explicitly model hidden constraints such as solver failures or infeasible geometries to improve search performance~\cite{archopt_hidden}. 
Also, \citet{bussemaker2021effectiveness} evaluated how well surrogate-assisted optimizers approximate the true Pareto front under fixed evaluation budgets, how efficiently they progress under convergence criteria, and how robust they remain in the presence of hidden constraints. 

Despite their success in other domains, surrogates remain comparatively underexplored in software engineering. 
Only a few works have attempted to introduce surrogate models in architecture optimization. 
The closest one is \cite{DBLP:conf/icsa/TitovP0H25}, which trained surrogates to approximate performance and modifiability metrics in tree-based architectural search. 
While the results demonstrate the feasibility of predicting non-functional values from architectural features, the design of the approach had practical limitations for ``live'' optimization scenarios. 
In particular, the surrogates were trained offline on precomputed datasets and had no influence on the optimization process.

In contrast to these earlier efforts, this work integrates surrogate models directly into the evolutionary architecture-refactoring algorithm. 
Our surrogates are trained incrementally on solver evaluations generated during search, and their predictions guide exploration by replacing a portion of costly solver invocations. 
This design allows the optimization to scale to larger design spaces while maintaining the quality of the solutions being explored.

\vspace{-0.5cm}\section{Approach}\label{sec:approach}\vspace{-0.3cm}

The multi-objective optimization process establishes a mapping between candidate architectural models and a multi-valued response surface, where each point represents the values of the non-functional properties of interest for an architecture. From a data-driven perspective, this process generates new pairs of architectural models and objective values, progressively forming a dataset that becomes the foundation for surrogate modeling. 


	\begin{figure}[t]
                \begin{subfigure}[b]{.46\textwidth}
                \includegraphics[width=\textwidth]{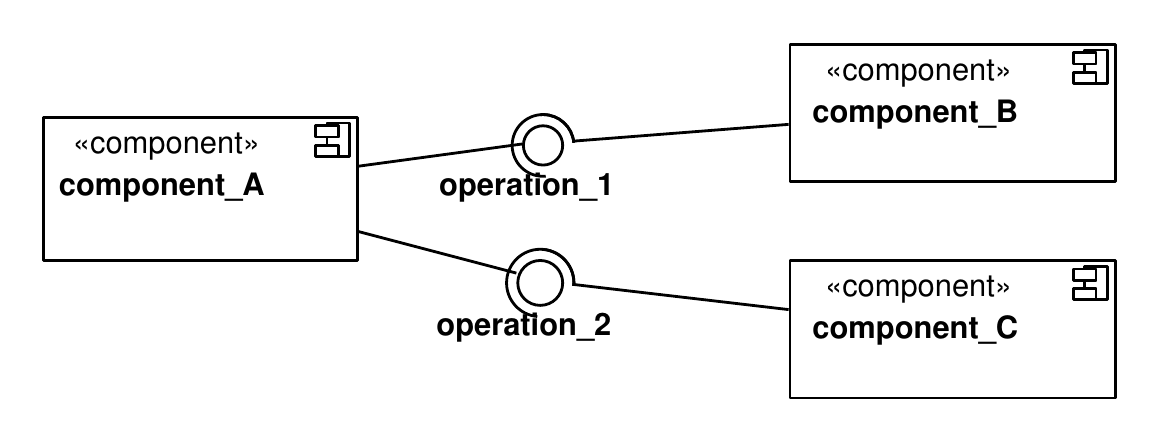}
                        \caption{Initial}\label{fig:mo2n-comp}
                \end{subfigure}
                \begin{subfigure}[b]{.46\textwidth}
                        \includegraphics[width=\textwidth]{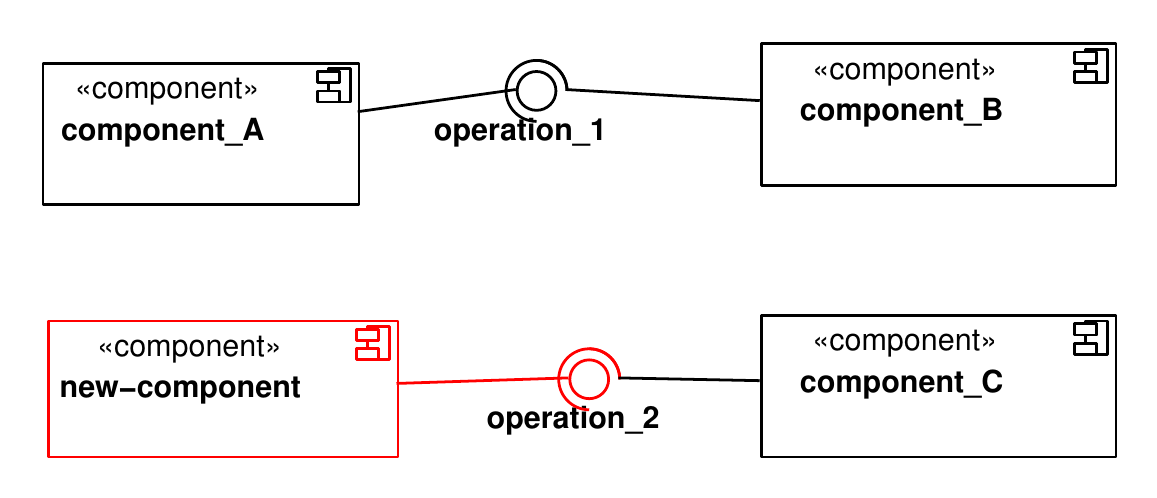}
                        \caption{Refactored}\label{fig:ref-mo2n-comp}
                \end{subfigure}
                \begin{subfigure}[b]{.46\textwidth}
                        \includegraphics[width=\textwidth]{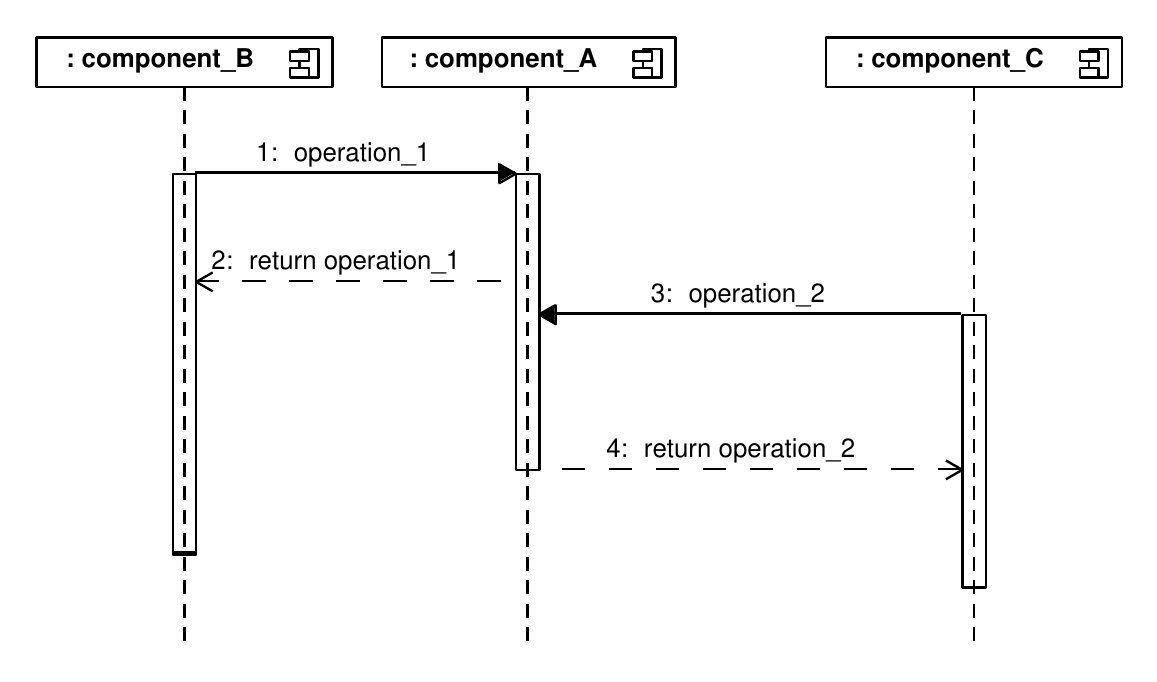}
                        \caption{Initial}\label{fig:mo2n-dynamic}
                \end{subfigure}
                \begin{subfigure}[b]{.46\textwidth}
                        \includegraphics[width=1.15\textwidth]{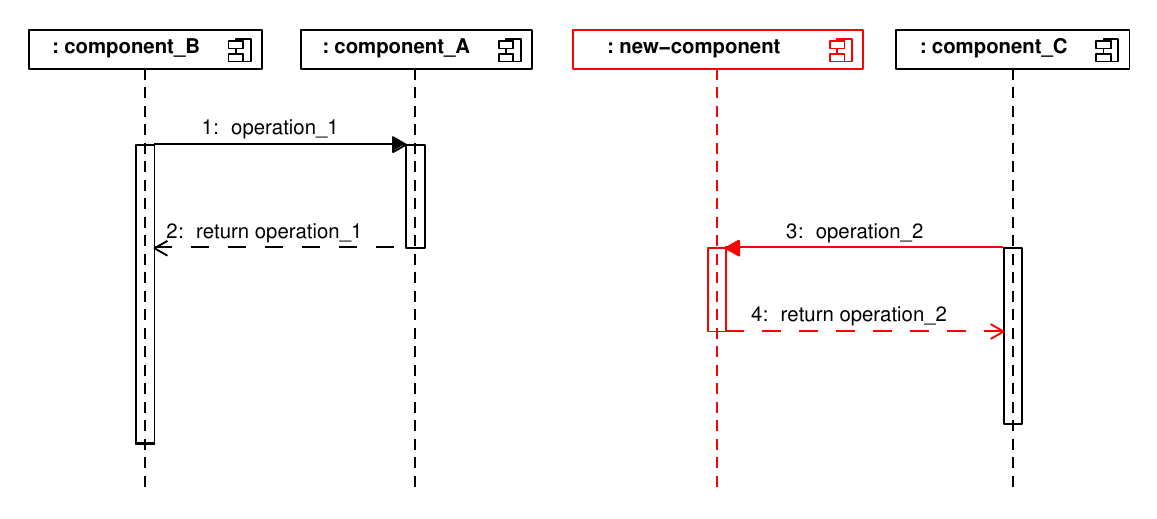}
                        \caption{Refactored}\label{fig:ref-mo2n-dynamic}
                \end{subfigure}
                \begin{subfigure}[b]{.46\textwidth}
                        \includegraphics[width=\textwidth]{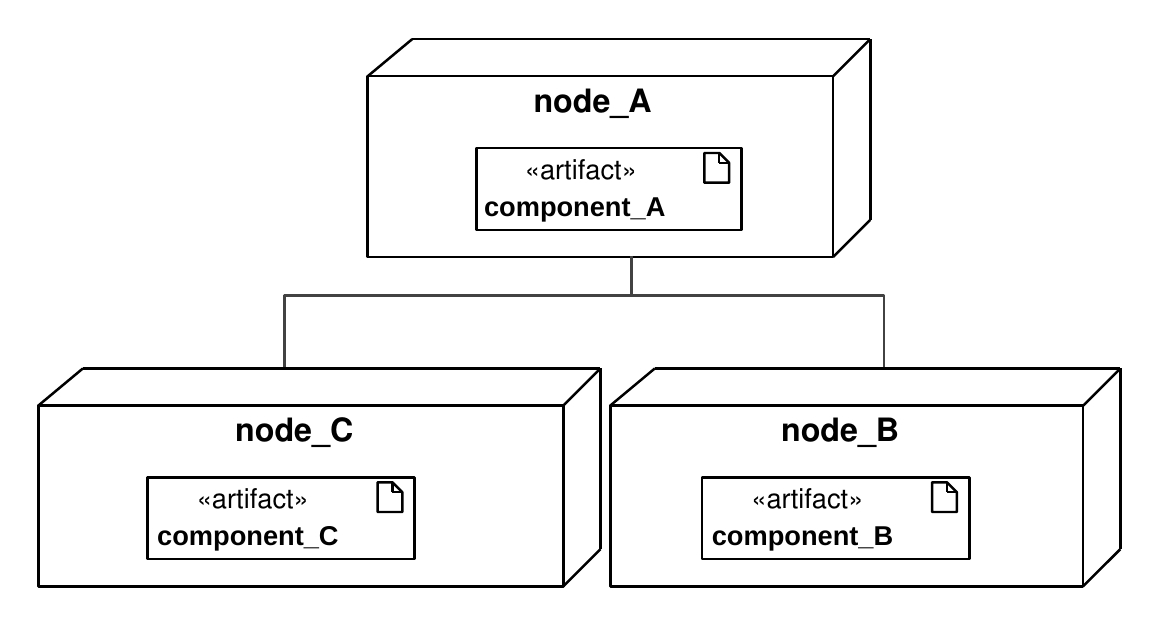}
                        \caption{Initial}\label{fig:mo2n-deploy}
                \end{subfigure}
                \begin{subfigure}[b]{.46\textwidth}
                        \includegraphics[width=\textwidth]{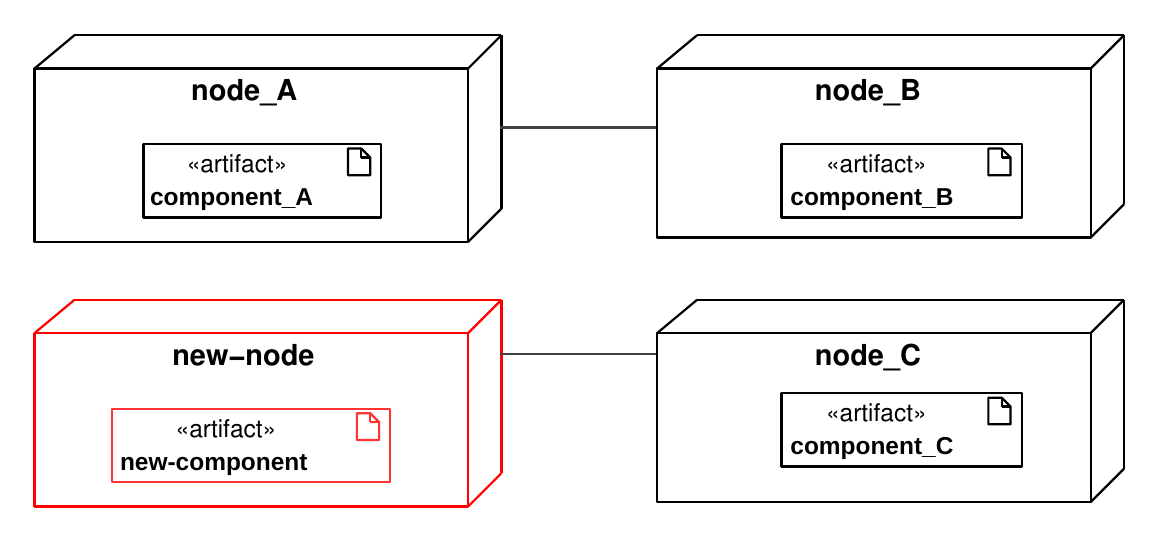}
                        \caption{Refactored}\label{fig:ref-mo2n-deploy}
                \end{subfigure}\vspace{-0.3cm}
                \caption{Example of action for selecting an operation to be moved to a new component to be deployed on a new node (\textit{motn}). The elements added by the refactoring action, across several UML views for the same architecture, are in red.}\vspace{-0.6cm}
                \label{fig:ref-mo2n-uml-diagrams}
        \end{figure}

Our work departs from an architectural refactoring optimization framework ~\cite{DBLP:journals/infsof/CortellessaPST23}, in which a genetic algorithm (GA) iteratively transforms architecture models using predefined refactoring actions and evaluates each candidate through solver-based analyses of non-functional properties.
\ADP{Any additional justification for choosing EASIER-kind of GA framework?}
\rev{The framework, called \texttt{EASIER}, is specifically designed for capturing architectural models using evolutionary algorithms and treating non-functional properties as objectives being evaluated on those models. 
Architectures are typically expressed as UML models, which are converted into quantitative models for analysis. These analysis models are usually supported by specialized solvers. NSGAII is the default GA supported by \texttt{EASIER} to find a set of Pareto-optimal architecture. Essentially, NSGAII involves the following steps: initialize a random population of candidate architectural models), evaluate fitness of individual architectures on multiple objectives (solver-based assessment of non-functional properties), perform a non-dominated sorting of these individuals, calculate a crowding distance for diversity, generate offspring via GA operators, and merge populations to select the best individuals. These steps are repeated for a fixed number of iterations. The GA operators are predefined refactoring actions for architectural models. For the non-functional properties, we consider competing objectives such as performance, reliability, energy and cost, each one with its corresponding analysis model.}

\rev{The refactoring catalog employed in this work comprises actions designed to enhance a variety of non-functional properties of a software system~\cite{DBLP:conf/icsa/CortellessaP024}. These actions involve: cloning a node (\texttt{clone}) to reduce utilization of an existing platform device, re-deploying a component to another node (\texttt{rede}) to reduce the load of the source node, transferring the logic of a given operation from one component to another (\texttt{move}), and selecting an operation to be moved to a new component to be deployed on a new node (\texttt{motn}) also to reduce load of the original component and node. An example of the \texttt{motn} action is given in \Cref{fig:ref-mo2n-uml-diagrams}}.
A feasibility engine~\cite{DBLP:conf/icsa/ArcelliCDP18} ensures the validity of the sequences of refactoring actions applied to the architectures.

\begin{figure}[t]
    \centering
    \includegraphics[width=0.8\linewidth]{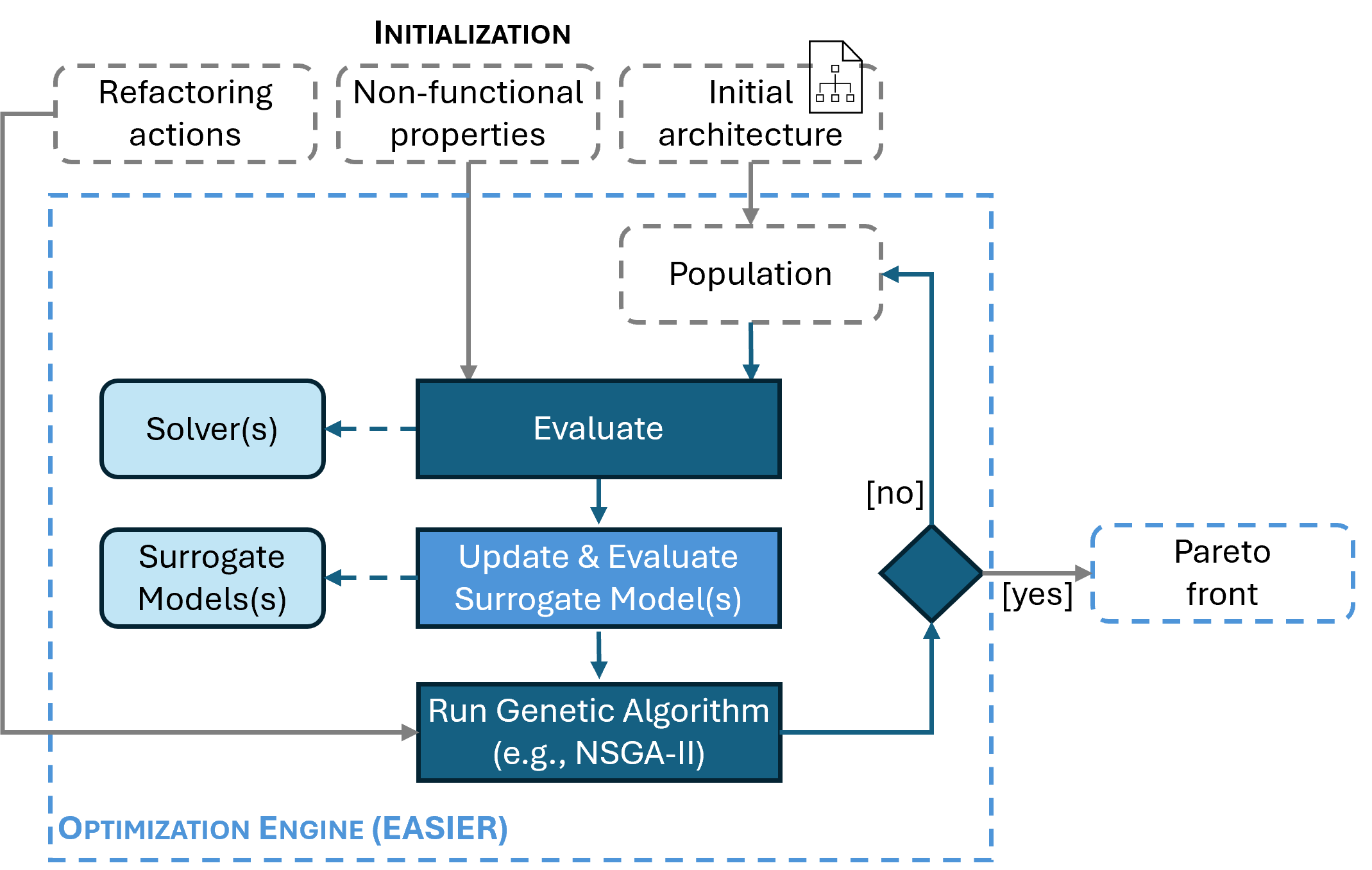}
    \caption{Workflow of the proposed approach integrating surrogate models into a genetic algorithm engine for architectural refactoring optimization (\texttt{EASIER}).}\label{fig:surrogate_workflow}\vspace{-0.4cm}
\end{figure}

We extend this base approach by integrating surrogate models (SMs) into the evaluation phase, as illustrated in \Cref{fig:surrogate_workflow}.
Our workflow begins with the initialization of a population of candidate solutions, each representing an architectural alternative of the software model. 
The fitness of each architecture with respect to a given objective can be computed either by invoking the corresponding solver, or by querying an SM. When an SM is used, the objective values are predicted from features extracted from the architecture. 
In each GA iteration, only a subset of the architectural population is evaluated using the solvers to obtain ground-truth objective values, while the remaining architectures are evaluated by the SMs. An initial set of surrogates is trained offline and injected into the optimization engine before the search begins. To provide flexibility, we maintain one SM per objective, allowing each surrogate to evolve independently based on the distribution and difficulty of its corresponding objective.

As SMs inevitably introduce prediction errors, and because the GA may explore regions of the design space that were not represented in the initial training data, the SMs can be incrementally updated through the search. At predefined iterations, their training sets are augmented with samples from the most recent solver evaluations, as proposed in \cite{DBLP:conf/icsa/TitovP0H25}. This periodic re-training helps maintain the relevance and accuracy of the surrogates during the optimization process.

The approach relies on UML to model the software architectures under optimization.
Specifically, as shown in \Cref{fig:ref-mo2n-uml-diagrams}, we utilize component diagrams to represent static relationships among software components, sequence diagrams to capture dynamic system behavior (\eg sequences of operations), and deployment diagrams to model the hardware architecture.
Since standard UML does not natively support performance modeling, we leverage established UML profiles~\cite{DBLP:conf/qest/LiAZCP17,DBLP:journals/infsof/CortellessaPST23,franks2008enhanced}.
For deployment nodes, we selected a set of representative Amazon EC2 instances\footnote{Amazon EC2 Instances Carbon Footprint Estimator: \url{https://docs.google.com/spreadsheets/d/1DqYgQnEDLQVQm5acMAhLgHLD8xXCG9BIrk-_Nv6jF3k}}, namely: \texttt{d2.2xlarge}, \texttt{m5ad.xlarge}, \texttt{t2.medium}, \texttt{t2.micro}, to capture diverse cost-performance trade-offs~\cite{DBLP:conf/icsa/CortellessaP024}.
Instance costs were taken from the Amazon EC2 Price History dataset\footnote{Amazon EC2 Spot Price History: \url{https://zenodo.org/doi/10.5281/zenodo.5880792}}.

\vspace{-0.5cm}\subsection{Feature Encoding of Architectural Models}\vspace{-0.3cm}

\begin{figure*}[t]
    \centering
    \includegraphics[width=0.99\linewidth]{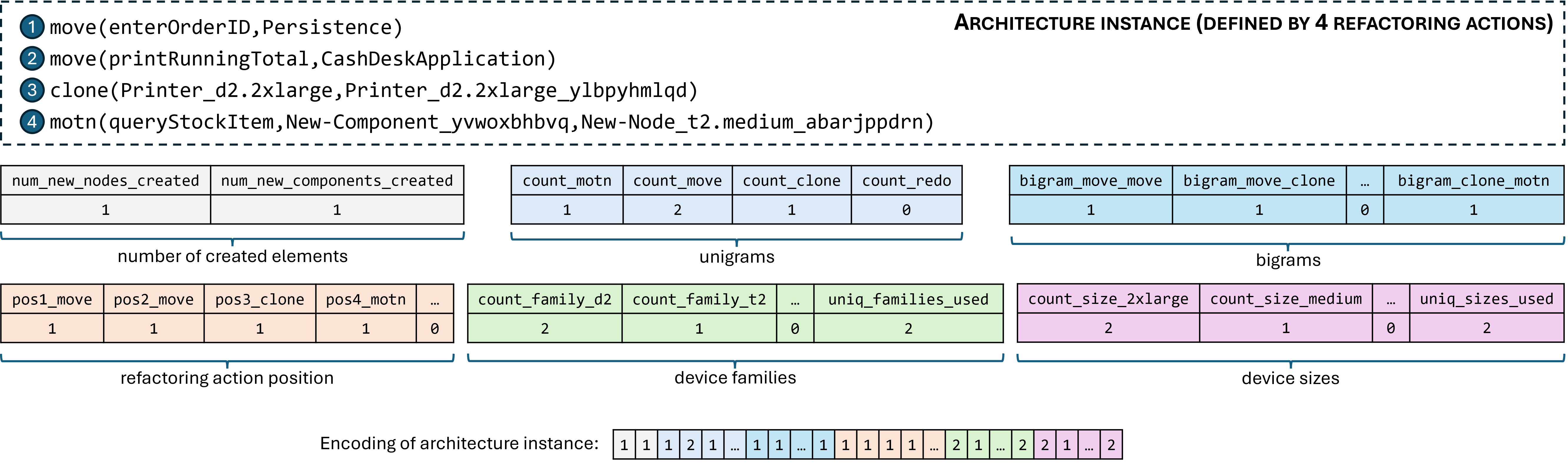}\vspace{-0.3cm}
    \caption{Encoding of a sequence of refactoring actions leading to a candidate architecture into feature vector with different feature categories. \ADP{Would it be possible to get the architecture representation for this sequence? Maybe only a partial view, such as the component view?}}\vspace{-0.6cm}
    \label{fig:feature_example}
\end{figure*}

Our SMs require a representation of each architecture of a population to a format (\ie features) suitable for regression. 
Since each architecture is defined by a sequence of refactoring actions (applied to an initial architecture), it cannot be processed directly by the SMs. 
Thus, we construct a tabular, numeric representation in which each candidate sequence is converted into a fixed-length feature vector describing its refactoring and architectural characteristics.
This encoding is defined once at the onset and remains stable in their features across all subsequent populations, considering that refactoring actions might introduce new nodes or components in some architectural models.

To encode the sequence structure in a way that is both compact and expressive, we borrow ideas from natural language processing \cite{jurafsky_speech_2024,wang2012baselines} and bio-molecular sequence analysis \cite{Leslie2001TheSK}. 
In those fields, sequences (e.g., words in a sentence or amino acids in a protein) are described not only by the individual elements that occur but also by the short-range patterns in which they appear. 
In our context, the encoding includes \textit{action unigrams}, which count how many times each refactoring action occurs, and \textit{action bigrams}, which count how often one action immediately follows another. 
While unigrams reflect the overall distribution of refactoring types, bigrams capture local dependencies that often carry architectural meaning. 
\rev{For example, a \texttt{clone} followed immediately by a \texttt{motn} suggests a distinct pattern that contributes to fault tolerance and load distribution, compared to the same actions appearing far apart in the sequence. }
By incorporating short-range co-occurrence patterns, we obtain a lightweight but informative description of how refactoring actions interact to lead to specific architecture configurations.
For instance, \Cref{fig:feature_example} shows how a sequence with four actions is mapped to a feature vector. 
Note that teatures are grouped into different categories, namely: elements created (nodes and components from UML models), action counts (unigrams), bigrams, positional action counts, and device families and sizes (for AWS characteristics of nodes), among others.

To retain information about the temporal structure of the transformation process, we additionally encode which actions appear at each position of the sequence. 
As some refactoring actions can introduce new architectural elements, the encoding also includes indicators quantifying the structural expansion of the architecture, such as the number of newly created components and nodes. 
For instance, having more components or nodes often correlates with horizontal scaling, functional decomposition, increased architectural complexity or potential points of failure, among other aspects. 

A representation challenge is that node identifiers in the architectural model often embed hardware information in non-uniform strings (\eg \texttt{StoreServer\-\_d2.2xlarge}) 
To retain this contextual information without relying on raw identifiers, we normalized node names by extracting two stable tokens: the hardware \textit{family} (\eg \texttt{t2}, \texttt{m6i}, \texttt{d2}) and the \textit{size} class (\eg \texttt{medium}, \texttt{xlarge}). 
These tokens are mapped to a finite vocabulary determined at initialization, and their occurrences are tracked across the refactoring actions. 

\vspace{-0.5cm}\subsection{Incremental Regression Models}\vspace{-0.2cm}

Once the architecture candidates are encoded, their non-functional objective values computed by the solvers 
are seen as regression targets for the SMs. 
Regression is a type of supervised learning in which the goal is to predict continuous values based on input data. 
Since we deal with multiple objectives, 
our approach internally creates a single regressor per objective.

We support incremental model training (see \Cref{fig:surrogate_workflow}) to reflect the dynamic nature of the GA optimization process. 
For the SMs, an initial regression model is learned using a fixed number of populations. The idea is to adjust the SM with new architecture candidates while still retaining patterns extracted from previous instances. 
Thus, every $k$ iterations, the model can be updated with a fraction of the new data available from each population. 
Gradient boosting techniques (e.g., \textit{XGBoost}) are appealing in this scenario, because the underlying estimators (tree ensembles) can be updated with newly-trained estimators for incoming data.

\vspace{-0.35cm}\section{Evaluation}\label{sec:results}\vspace{-0.35cm}

\rev{The  research methodology is an empirical study based on a well-known case-study from the literature, on which we performed different experiments.}
We evaluate our approach by formulating two research questions that guided our experimental design. 
We aim to assess how surrogate models can support an efficient multi-objective optimization of architectural models.

\smallskip

\noindent\textbullet~\textbf{RQ1:} \textit{Can surrogate models reduce the overall execution time of multi-objective architectural refactoring optimization while maintaining solution quality?}

This question investigates whether integrating SMs into the optimization process yields meaningful reductions in execution time compared to exclusively relying on solver-based analyses. Execution time was selected the definitive metric to ensure findings remain independent of specific hardware resource allocations and the fluctuating economics of cloud environments, offering a direct picture of the optimization's efficiency.
Furthermore, the question examines the effect of surrogate models on the quality of the Pareto solutions, using indicators such as hypervolume and inverted generational distance to determine how closely the obtained front approximates the reference one.

\smallskip

\noindent\textbullet~\textbf{RQ2:} \textit{Are there differences in the structural characteristics of the architecture candidates generated by a surrogate-assisted approach and those generated by a standard search-based optimization engine?}

Unlike the previous question that looks at the objective space, this question focuses on the architectural models being explored during the optimization process, and how surrogate models shape the use of certain refactoring actions for the architecture candidates.

\vspace{-0.4cm}\subsection{Experimental Setup}\vspace{-0.2cm}

We compared a standard multi-objective architecture optimization setup using \textit{NSGA-II} as the underlying GA with our surrogate-assisted approach, in which regression models approximate solver evaluations\footnote{We chose to use NSGA-II because it is widely used in the literature. Recently, NSGA-III has shown advantages over NSGA-II only with a large set of objectives ($15+$)~\cite{DBLP:journals/tec/DebJ14}.}.
The surrogates were implemented using \textit{XGBoost}, which supports incremental model updates. The regression models were generated and consumed by the optimization engine using a FastAPI server.
Once pre-trained, the SMs provided predictions of objectives values for each population. In each optimization iteration, $50\%$ of the candidate architectures were evaluated with the solver, while the remaining architectures were assigned objective values predicted by the SMs. The SMs were re-trained at pre-specified iterations

Regarding the SMs, we used two strategies: a pre-trained model (without re-training), and a periodic re-training mode in which the models were re-trained every $k$ steps.
The former strategy was used as a baseline for estimating the execution time reduction of SMs, while the latter served to evaluate the potential benefits of updating SMs during the optimization process.
To evaluate the impact of re-training SMs, two frequencies ($k$ step) were tested: every $2$ and $5$ iterations. 

Experiments were conducted on a well-known case study from the software architecture optimization literature. 
We evaluated performance along two dimensions: computational time and quality indicators for the resulting Pareto front (\textbf{RQ1}). 
We also analyzed the structural characteristics of the architecture candidates generated by both approaches, focusing on the refactoring actions applied to the architectural models (\textbf{RQ2}).

The GA configuration followed common literature settings~\cite{DBLP:conf/icsa/ZhaoMB25,DBLP:conf/icsa/ArcelliCDP18,DBLP:journals/infsof/NiDYMYHX21}: a population of $12$ architecture candidates, chromosomes (\ie a sequence of refactoring actions) of length $4$, and $102$ iterations per run. Each configuration was executed $31$ times on each system to mitigate variability\footnote{These parameters reflect typical choices in architecture optimization studies, considering that each population member represents a full software architecture model and each evaluation involves running computationally intensive analysis tools.}.
\rev{Furthermore, the optimization process was configured to evaluate $7$ objectives, involving performance, reliability, energy, price and number of changes, among others. Therefore, the found Pareto fronts include solutions, i.e., alternative architectures, that represent different trade-offs among these objectives.}

For deployment nodes, we selected a set of representative Amazon EC2 instances\footnote{Amazon EC2 Instances Carbon Footprint Estimator: \url{https://docs.google.com/spreadsheets/d/1DqYgQnEDLQVQm5acMAhLgHLD8xXCG9BIrk-_Nv6jF3k}}, namely: \texttt{d2.2xlarge}, \texttt{m5ad.xlarge}, \texttt{t2.medium}, \texttt{t2.micro}, to capture diverse cost-performance trade-offs~\cite{DBLP:conf/icsa/CortellessaP024}.
Instance costs were taken from the Amazon EC2 Price History dataset\footnote{Amazon EC2 Spot Price History: \url{https://zenodo.org/doi/10.5281/zenodo.5880792}}.
\rev{It is worth noting that we selected the AWS instances based on the availability of datasets, which to the best of our knowledge, are the only ones that provide detailed information on both performance and energy consumption for a variety of instance types\footnote{These values can easily changed whether other datasets become available. More important, using these AWS datasets do not limit generalizability of the approach.}.} 

Experiments were carried out on a cluster of $3$ Dell PowerEdge C6525 servers, each equipped with $2$ AMD EPYC 7282 2.80GHz CPUs and 512 GiB of RAM\footnote{The replication package is available at \url{https://doi.org/10.5281/zenodo.17846462}}.
The target case-study was \emph{CoCoME (CCM)}, which is a reference system for non-functional model-based analyses~\cite{Herold-Klus-Welsch-Deiters-Rausch-Reussner-Krogmann-Koziolek-Mirandola-Hummel}. 
It describes a trading system consisting of several stores, each containing one or more cash desks. 
A cash desk is equipped with hardware and software elements required to serve customers (\eg a cash box, printer, bar bode scanner). 
\emph{CCM} captures operational scenarios such as scanning products, processing payments, and managing stock replenishment.


\rev{For the initial training of the surrogates (one per objective), we took populations from two \textit{EASIER} iterations ($\approx 32$ architectures). Although this is a small sample size, the SMs can be adjusted quickly with periodic updates from the optimization process. \textit{XGBoost} was configured with $500$ estimators and a learning rate of $0.1$. This pretraining setting is not computationally expensive ($3$ min on average in total) for \textit{XGBoost}.}

\vspace{-0.35cm}\subsection{Metrics}\label{sec:approach:quality_indicators}\vspace{-0.4cm}

To evaluate the quality of the Pareto fronts produced by the optimization approaches, we rely on standard multi-objective quality indicators. In particular, we employ Hypervolume (HV)~\cite{Cao-Smucker-Robinson-2015,Beume-Naujoks-Emmerich-2007}, Inverted Generational Distance (IGD+)~\cite{Ishibuchi-Masuda-Tanigaki-Nojima-2015}.
These indicators were selected due to  their prevalence in prior literature and their complementary nature \cite{Li-Yao-2020}.
For the \texttt{IGD+} indicator that requires a reference front, we constructed it by merging the non-dominant solutions from all runs. 
For \texttt{HV} that requires a nadir point, such a reference point was defined as the worst value observed per objective across all executions.\vspace{0.15cm}

\vspace{-0.5cm}\section{Results}\vspace{-0.35cm}

\subsection{RQ1: Reductions in overall execution time}\vspace{-0.2cm}

\begin{figure}[t]
        \centering
        \includegraphics[width=\linewidth]{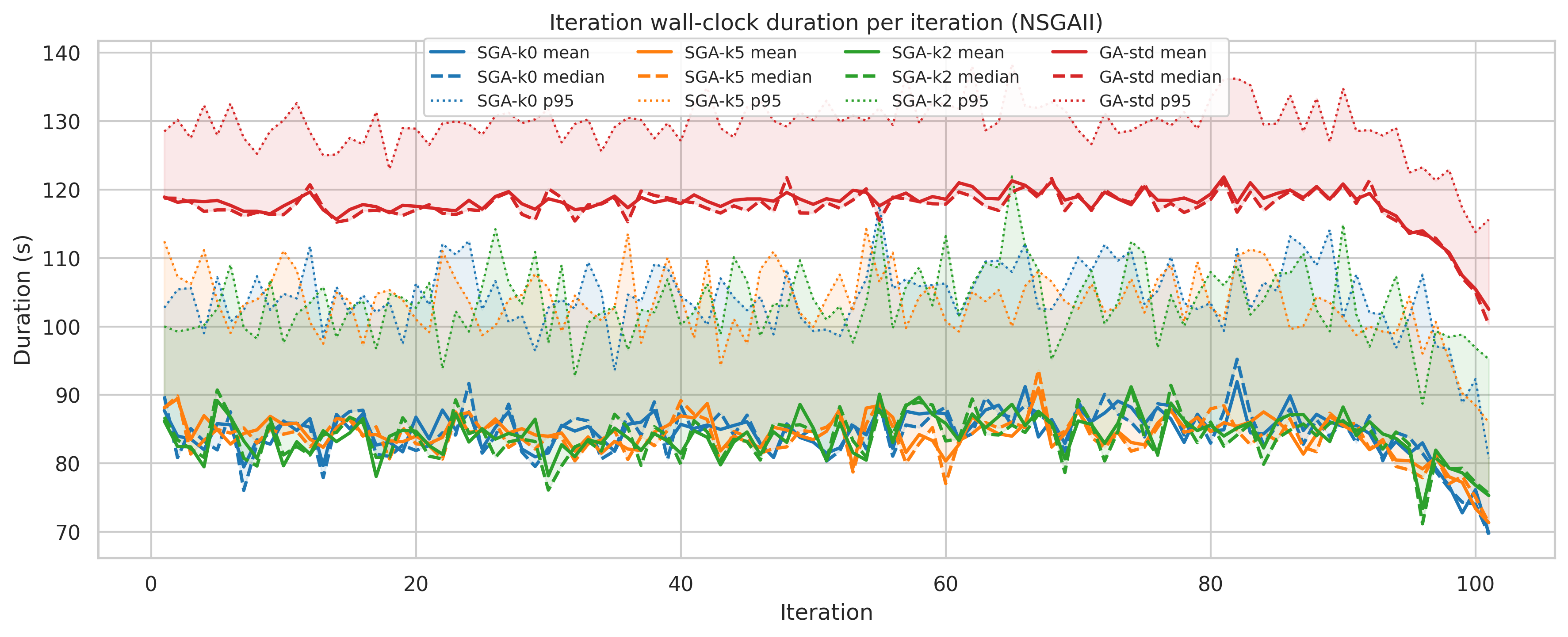}\vspace{-0.3cm}
    \caption{Per-iteration execution time of the standard GA and the surrogate-based GA across both case studies. \emph{GA-std}: plain genetic algorithm without surrogate. \emph{SGA-k0}: pre-trained surrogate model (no re-training). \emph{SGA-k2}, \emph{SGA-k5}: surrogate re-trained every $2$ and $5$ iterations, respectively.}\vspace{-0.3cm}\label{fig:ccm:execution_time}
        \label{fig:ccm:execution_time}
\end{figure}

A comparison of execution times between standard GA and surrogate-based GA is presented in \Cref{fig:ccm:execution_time}.
\rev{We remark that we have considered the execution time as primary cost metric for our evaluation, as it is a direct measure of the computational efficiency of the optimization process. Furthermore, the execution time is a critical factor in determining the practical applicability of evolutionary-based optimization methods in real-world scenarios.}
The standard GA exhibited the highest execution time in the first iteration, which then stabilized as iterations progressed.
This behavior was expected due to various optimization steps that occurred throughout the optimization process.
In contrast, the surrogate-based GA did not exhibit that behavior and achieved a substantial speedup across all iterations, reducing the overall execution time by $\approx 30\%$. 


\begin{table}[t]
\centering
\caption{Pairwise comparisons between GA-std and SGA variants using Wilcoxon signed-rank tests with Holm correction. \emph{$p_{\text{Holm}}$}: Holm-corrected $p$-value; \emph{$d_z$}: Cohen's effect size. \emph{GA-std}: plain genetic algorithm without surrogate. \emph{SGA-k0}: pre-trained surrogate model (no re-training). \emph{SGA-k2}, \emph{SGA-k5}: surrogate re-trained every $2$ and $5$ iterations, respectively.}\vspace{0.1cm}
\begin{tabular}{l | c c c | c c c}
\toprule
Comparison & Indicator & $p_{\text{Holm}}$ & $d_z$ & Indicator & $p_{\text{Holm}}$ & $d_z$\\
\midrule
GA-std vs SGA-k0 & HV   & 0.289 & -0.28 & IGD+ & 1.000 & -0.19\\
GA-std vs SGA-k2 & HV   & 0.546 & -0.20 & IGD+ & 1.000 &  0.01\\
GA-std vs SGA-k5 & HV   & 0.193 & -0.44 & IGD+ & 0.471 & -0.30\\

\bottomrule
\end{tabular}
\vspace{-0.7cm}
\label{tab:stats}
\end{table}

As expected\footnote{The quality indicators are expected to be better for the standard genetic algorithm.}, GA-std achieves the best median performance for both indicators across 31 independent runs. 
In particular, GA-std attains the highest HV, indicating superior coverage of the objective space. 
Similarly, GA-std yields the lowest IGD+ value, suggesting better proximity to the reference Pareto front and improved distribution of solutions.
Among the SGA variants, SGA-k2 shows the closest performance to GA-std, while SGA-k5 consistently exhibits lower HV and higher IGD+ values, indicating reduced Pareto front quality.

\ADP{Please double-check if this justification is Ok?}

\rev{For the comparisons,  the variants considered the same group under different scenarios (the quality indicators). Since data was not normally distributed, we used a Friedman test to check for differences and Wilcoxon signed-rank as a post-hoc test to identify differences. }The Friedman test revealed a significant difference among variants for HV ($\chi^2 = 9.50$, $p = 0.023$), but not for IGD+ ($p = 0.218$). Pairwise comparisons between GA-std and SGA variants using Wilcoxon with Holm correction (due to repeated measures) are reported in \Cref{tab:stats}.
No pairwise comparison reached statistical significance after correction ($\alpha = 0.05$). 
However, effect sizes provide additional insights. 
For HV, GA-std shows a moderate advantage over SGA-k5 ($d_z = -0.44$) and smaller effects over SGA-k0 and SGA-k2. 
For IGD+, effect sizes are small, with a moderate trend favoring GA-std over SGA-k5 ($d_z = -0.30$).

The results indicate that GA-std consistently outperforms SGA variants in terms of both coverage (HV) and proximity/diversity (IGD+). 
While these differences are not statistically significant under multiple comparison correction, the observed effect sizes suggest that the performance gap{|}particularly between GA-std and SGA-k5{|}is practically meaningful.
Overall, the SGA variants do not provide improvements over the standard GA in this setting, and in some configurations (\eg SGA-k5) may lead to a degradation in Pareto front quality.

\paragraph*{Summary.}

Answering \textbf{RQ1}, the surrogate-assisted approach provides substantial runtime savings by reducing expensive solver calls, while remaining competitive in solution quality when re-training is frequent.
Thus, surrogates with frequent re-training can preserve quality of Pareto fronts despite relying on approximate evaluations.
Furthermore, quality indicators indicate that differences in Pareto-front quality are limited: \texttt{HV} shows only a global trend without significant Holm-corrected pairwise contrasts, and \texttt{IGD+} shows no significant differences.
Overall, \texttt{GA-std} remains a strong baseline, but \texttt{SGA-k2} achieves comparable quality, especially for \texttt{IGD+}. 
This aspect is relevant because surrogates reduced execution time by $\approx 30\%$, highlighting a practical quality-time tradeoff.


\vspace{-0.4cm}\subsection{RQ2: Characterization of architectural models}\vspace{-0.2cm}

\begin{figure}[t]
    \centering
    \includegraphics[width=0.65\linewidth]{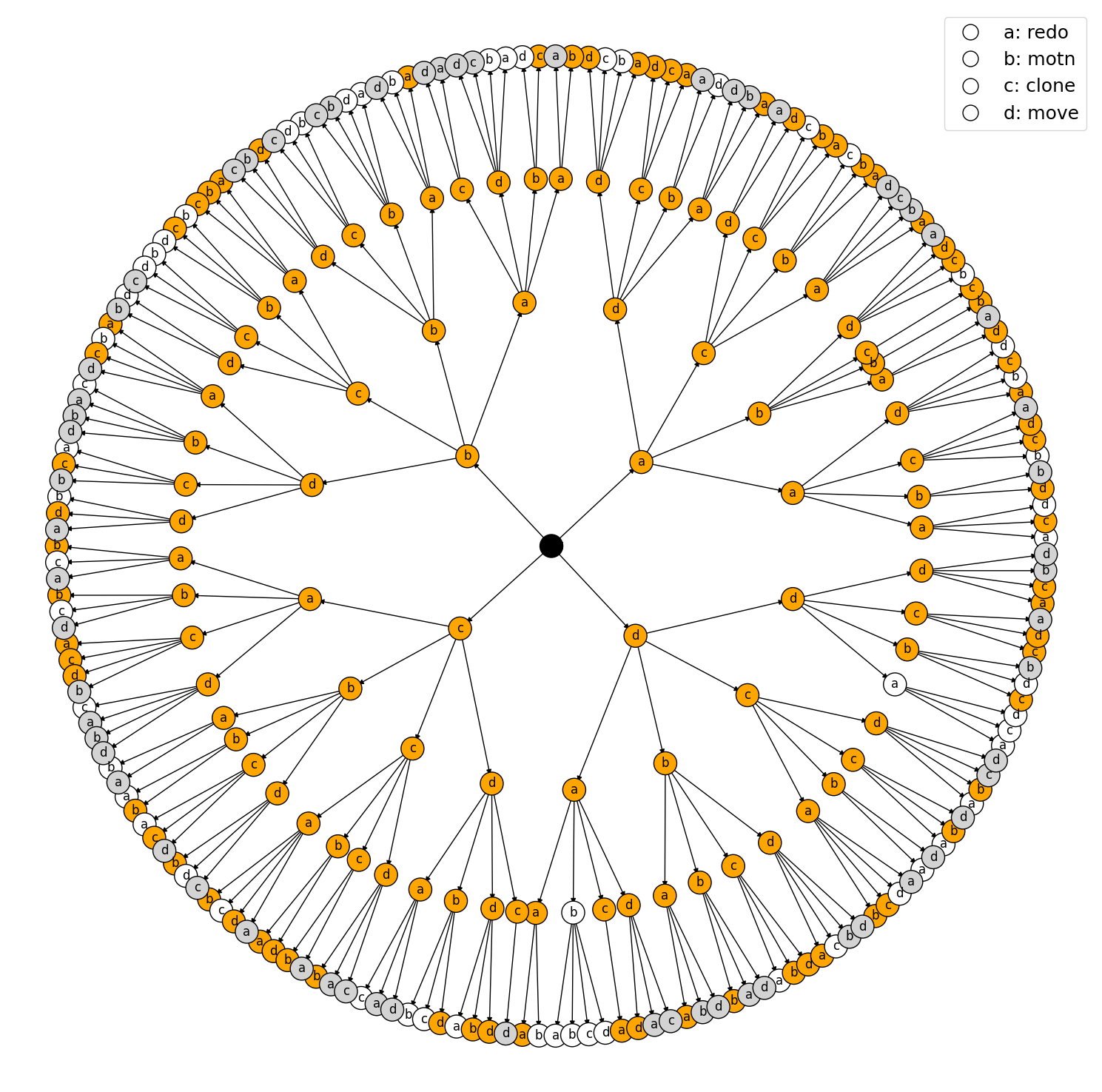}\vspace{-0.3cm}
    \caption{Sequences of refactoring actions for architectures in the Pareto fronts of both GAs. White nodes indicate sequences only generated by the standard GA, while gray nodes indicate sequences only generated by the surrogate-assisted GA ($k=5$). Yellow nodes correspond to intersecting sequences from algorithms. \ADP {Probably reduce or remove figure from paper to gain space}}\vspace{-0.75cm}
    \label{fig:ccm-prefix-tree-comparison}
\end{figure}

To analyze differences in the architectural models between the standard and surrogate-assisted GAs, we used the refactoring actions, described in \Cref{sec:approach}, as proxies for understanding the corresponding software architecture. 

For \emph{CCM}, we considered all the sequences of refactoring actions of the architectures observed in the Pareto fronts of both GAs.
For simplicity, we omitted the parameters of the actions and retained only their types. The action sequences were all arranged in a (prefix) tree, as schematially shown in \Cref{fig:ccm-prefix-tree-comparison}. A significant intersection of sequences was observed and, interestingly, the main differences between the two algorithms lie in the last refactoring action of the sequences. This trend was similar, regardless of the \textit{k} step (for re-training) chosen for the surrogate models.

While the standard GA employed actions \texttt{motn} (move operation to  new component on new node) and \texttt{clone} (clone node) more frequently, the surrogate-assisted GA relied more on actions \texttt{redo} (re-deploy existing component) and \texttt{move} (relocate operation to existing node). The different action usage profiles indicate differences in the resulting architectural models, and it can also affect the achievement of particular non-functional objectives. For example, since the \texttt{motn} and \texttt{clone} actions tend to expand the system infrastructure, the standard GA may favor solutions with scalability and robustness (probably involving a higher resource consumption). On the other hand, the preference of the surrogate-assisted GA for actions that re-distribute system components suggests changes that lead to efficient and less costly solutions. As a result, the SMs could contribute to architectures that better satisfy energy, power or price objectives, but lower gains for performance or reliability objectives, when compared to a pure GA algorithm. It should be noticed that these architectural aspects are difficult to assess when considering only quality indicators for the Pareto front.

\bigskip

\vspace{-0.7cm}\paragraph*{Summary.} 
Our evidence for answering \textbf{RQ2} shows that 
a significant portion of the sequences explored by both GAs are shared, which means that the surrogate models do not drastically alter the search space exploration in terms of architectural modifications. Nevertheless, the variations in the usage of refactoring actions (towards the end of the sequences) tell that the two types of GAs seem to explore slightly different areas of the design space. Furthermore, these action profiles can influence the achievement of specific non-functional objectives.

\vspace{-0.5cm}\subsection{Lessons Learned}\vspace{-0.2cm}\label{sec:lessons_learned}
Our results highlight a practical trade-off: surrogate models reduce execution time by decreasing expensive invocations to analysis solvers, while Pareto-front quality remains broadly comparable but can vary across indicators (\textbf{RQ1}).
At the same time, standard and surrogate-assisted GAs exhibit relatively similar patterns of refactoring-action usage across sequence positions, except for the last action of the sequence, consistent with the findings of \textbf{RQ2}. The architectural solutions explored by the surrogate-assisted GA are not necessarily a subset of those explored by a standard GA. Although more experimentation is needed, our results suggest that SMs can lead to solutions with alternative tradeoffs.

\rev{The predictive performance of the SMs depend mainly on having a more or less regular (e.g., normal) distribution of the series of objective values, which makes them treatable with non-linear regression. For \textit{XGBoost}, in particular, we were able to obtain good results with a relatively small number of estimators, which made (re-)training quite fast. Having a fixed set of features also contributed to keeping the execution times low, because it makes the feature set independent from the sizes of the architectural models. Increasing the number of estimators, e.g., to achieve better predictive performance or to compensate for irregular objective distributions, would likely increase the SM training times.} 

\rev{As for the encoding strategy, it proved to be appropriate for the \textit{CoCoME} predictions for different architectural models, by relying on unigrams and bigrams about components, nodes and refactoring actions, regardless of the size or structural variations of the architectural models. Based on initial results with another case study (\texttt{TrainTicket}), we believe that the encoding strategy can be successfully extrapolated to other systems.}

Overall, surrogate-assisted optimization is promising, especially under constrained computational budgets, while motivating further study of surrogate accuracy and role of action usage profiles in the search through the design space.


\vspace{-0.5cm}\section{Threats to validity}\label{sec:t2v}\vspace{-0.2cm}
We identified several threats to internal, construct, and external, and conclusion validity in our experimental evaluation.

\vspace{-0.3cm}\paragraph{Construct validity.}
A threat arises from the lack of feature selection or feature importance analysis, 
as we included all features from our encoding schema without filtering them. 
Irrelevant or redundant features might have introduced noise into the regression models, while key features might not have been emphasized appropriately. 
In addition, we split the population instances into halves to decide which ones were tackled with the solvers and which ones were predicted with the SMs, but we did not try alternative sampling ratios.
These factors could have affected the predictive accuracy of surrogates or their training times.

\vspace{-0.3cm}\paragraph{External validity.}
The generalizability of our results is limited by several experimental assumptions. 
Our evaluation relies on architectural models expressed in a particular modeling notation.
Although UML notation is widely used in performance and reliability analysis, other modeling languages might allow different architectural structures, refactoring primitives, or analytic solvers. 
As a result, the behavior of SMs might vary when applied to another setting. 

A further threat arises from the limited refactoring space used by the optimization engine. Architecture candidates were generated using a predefined set of four refactoring actions and a fixed sequence length, which stem from common architecture refactorings. 
Finally, the datasets used are relatively small compared to typical machine learning benchmarks. Surrogate behavior, accuracy trends, and sampling effects may differ in larger or noisier datasets, especially those produced by long-running or industrial-scale optimization processes. 

\vspace{-0.3cm}\paragraph{Internal validity.}
Parts of our machine learning pipeline involve inherent non-determinism. Although we controlled randomness through repeated runs and fixed seeds, different configurations and executions might lead to variations in training/test splits, sampling strategy or surrogate performance, potentially affecting our conclusions. 
Also, differences in the execution environment (\eg hardware, engine configuration) might affect timing measurements.

\vspace{-0.3cm}\paragraph{Conclusion validity.}

We did not evaluate human factors such as how architects interpret or trust surrogate predictions and the derived architectural models, which might influence the practical adoption of surrogate-based optimization.

\vspace{-0.5cm}\section{Conclusion}\label{sec:conclusion}\vspace{-0.35cm}

In this work, we investigated the feasibility of using surrogate models to accelerate multi-objective quality-attribute optimization of architectural models. 
The central idea was to interleave expensive solver-based evaluations with fast, but potentially less accurate, surrogate predictions, thereby reducing computation time while preserving the effectiveness of the optimization process. 
Our study confirms that surrogate models can be integrated into the architecture optimization process with minimal overhead and that they offer meaningful efficiency gains in realistic settings, albeit with less variability on the Pareto fronts. \rev{The execution time reductions are important, as the computational cost is a typical concern for multi-objective optimization, and also because in the architecture domain these reductions can enable practitioners to explore larger design spaces (with a given budget).}
Furthermore, the search space (\ie the structure of the architecture candidates) explored by the surrogate-assisted GA seems to differ slightly from that taken as the baseline. More experiments should be performed to confirm this trend and the effects of surrogates on the objective values. \rev{Also, we would like to study whether surrogates can have a potential impact on the monetary cost and resource consumption of the optimizations.}

We envision several directions for future work for our approach. Incorporating drift detection over incoming data batches would make it possible to selectively re-train the surrogates upon data distribution changes, thus improving the performance of the current approach. Exploring  alternative encodings for architectural models and refactoring actions (\eg graph-based or embeddings techniques) may yield more informative feature spaces for surrogate modeling. 
Finally, the decision of which candidates to evaluate with solvers can be enhanced through adaptive sampling strategies that account for uncertainty, diversity, or predicted model reliability.

\vspace{-0.55cm}\subsubsection{Data Availability.} For the surrogate implementation, the datasets, and experimental results see  \url{https://github.com/danieledipompeo/easier-surrogate}.

\vspace{-0.5cm}\subsubsection{Acknowledgments.} This research was funded in whole or in part by the Austrian Science Fund (FWF): \href{https://doi.org/10.55776/COE12}{10.55776/COE12}. Also, this work was partially supported by PICT-2021-00757 project (Argentina).




\vspace{-0.3cm}\bibliography{biblio}

@inproceedings{DBLP:conf/icsa/ArcelliCDP18,
  author       = {Davide Arcelli and
                  Vittorio Cortellessa and
                  Mattia D'Emidio and
                  Daniele {Di Pompeo}},
  title        = {{EASIER:} An Evolutionary Approach for Multi-objective Software ArchItecturE
                  Refactoring},
  booktitle    = {18th {IEEE} International Conference on Software Architecture, {ICSA}},
  pages        = {105--114},
  __publisher    = {{IEEE} Computer Society},
  year         = {2018},
  _url          = {https://_doi.org/10.1109/ICSA.2018.00020},
  _doi          = {10.1109/ICSA.2018.00020}
}

@inproceedings{DBLP:conf/icsa/ZhaoMB25,
  author       = {Yiming Zhao and
                  Tiziano De Matteis and
                  Justus Bogner},
  title        = {How Does Microservice Granularity Impact Energy Consumption and Performance?
                  {A} Controlled Experiment},
  booktitle    = {22nd {IEEE} International Conference on Software Architecture, {ICSA}},
  pages        = {84--95},
  __publisher    = {{IEEE}},
  year         = {2025},
  _url          = {https://_doi.org/10.1109/ICSA65012.2025.00018},
  _doi          = {10.1109/ICSA65012.2025.00018}
}

@article{diaz-paceRoleSearchBudgets2025,
  title = {On the Role of Search Budgets in Model-Based Software Refactoring Optimization},
  author = {Diaz-Pace, J.~Andres and {Di Pompeo}, Daniele and Tucci, Michele},
  journaltitle = {Automated Software Engineering},
  volume = {33},
  number = {1},
  pages = {18},
  issn = {1573-7535},
  year = {2025},
  _doi = {10.1007/s10515-025-00564-y},
  _url = {https://_doi.org/10.1007/s10515-025-00564-y}
}

@article{DBLP:journals/infsof/NiDYMYHX21,
  author       = {Youcong Ni and
                  Xin Du and
                  Peng Ye and
                  Leandro L. Minku and
                  Xin Yao and
                  Mark Harman and
                  Ruliang Xiao},
  title        = {Multi-objective software performance optimisation at the architecture
                  level using randomised search rules},
  journal      = {Inf. Softw. Technol.},
  volume       = {135},
  pages        = {106565},
  year         = {2021},
  _url          = {https://_doi.org/10.1016/j.infsof.2021.106565},
  _doi          = {10.1016/J.INFSOF.2021.106565}
}

@inproceedings{DBLP:conf/icsa/TitovP0H25,
  author       = {Vadim Titov and
                  Jorge Andr{\'{e}}s D{\'{\i}}az Pace and
                  Sebastian Frank and
                  Andr{\'{e}} van Hoorn},
  title        = {Architecture Optimization using Surrogate-based Incremental Learning
                  for Quality-attribute Analyses},
  booktitle    = {22nd {IEEE} International Conference on Software Architecture, {ICSA}},
  pages        = {278--288},
  _publisher    = {{IEEE}},
  year         = {2025},
  _url          = {https://_doi.org/10.1109/ICSA65012.2025.00035},
  _doi          = {10.1109/ICSA65012.2025.00035}
}

@inbook{Arcuri-Fraser-2011,
        title        = {On Parameter Tuning in Search Based Software Engineering},
        author       = {Arcuri, Andrea and Fraser, Gordon},
        year         = 2011,
        booktitle    = {Search Based Software Engineering},
        _publisher    = {Springer Berlin Heidelberg},
        _address      = {Berlin, Heidelberg},
        series       = {LNCS},
        volume       = 6956,
        pages        = {33–47},
        _doi          = {10.1007/978-3-642-23716-4_6},
        _isbn         = {978-3-642-23715-7},
        _url          = {http://link.springer.com/10.1007/978-3-642-23716-4_6},
        editor       = {Cohen, Myra B. and Ó Cinnéide, Mel},
        collection   = {LNCS}
}

@article{Luong-Nguyen-Gupta-Rana-Venkatesh-2021, 
        title={Adaptive cost-aware Bayesian optimization}, 
        volume={232}, 
        ISSN={0950-7051},
        _doi={10.1016/j.knosys.2021.107481},
        journal={Knowl. Based Syst.},
        author={Luong, Phuc and Nguyen, Dang and Gupta, Sunil and Rana, Santu and Venkatesh, Svetha},
        year={2021},
        month=11,
        pages={107481}
}

@inbook{Meedeniya-Buhnova-Aleti-Grunske-2010,
        title        = {Architecture-Driven Reliability and Energy Optimization for Complex Embedded Systems},
        author       = {Meedeniya, Indika and Buhnova, Barbora and Aleti, Aldeida and Grunske, Lars},
        year         = 2010,
        booktitle    = {Research into Practice – Reality and Gaps},
        _publisher    = {Springer Berlin Heidelberg},
        _address      = {Berlin, Heidelberg},
        series       = {LNCS},
        volume       = 6093,
        pages        = {52–67},
        _doi          = {10.1007/978-3-642-13821-8_6},
        _isbn         = {978-3-642-13820-1},
        _url          = {http://link.springer.com/10.1007/978-3-642-13821-8_6},
        editor       = {Heineman, George T. and Kofron, Jan and Plasil, Frantisek},
        collection   = {LNCS}
}

@inproceedings{Martens-Koziolek-Becker-Reussner-2010,
        title        = {Automatically improve software architecture models for performance, reliability, and cost using evolutionary algorithms},
        author       = {Martens, Anne and Koziolek, Heiko and Becker, Steffen and Reussner, Ralf},
        year         = 2010,
        _month        = {Jan},
        booktitle    = {Proceedings of the First Joint WOSP/SIPEW Int. Conf. Perform. Eng.},
        _publisher    = {ACM},
        _address      = {San Jose California USA},
        pages        = {105–116},
        _doi          = {10.1145/1712605.1712624},
        _isbn         = {978-1-60558-563-5},
        _url          = {https://dl.acm.org/_doi/10.1145/1712605.1712624},
        language     = {en}
}

@inproceedings{Rago-Vidal-Diaz-Pace-Frank-van-Hoorn-2017,
        title        = {Distributed quality-attribute optimization of software architectures},
        author       = {Rago, Alejandro and Vidal, Santiago and Diaz-Pace, J. Andres and Frank, Sebastian and van Hoorn, Andr{\'{e}}},
        year         = 2017,
        _month        = {Sep},
        booktitle    = {Proceedings of the 11th SBCARS},
        _publisher    = {ACM},
        _address      = {Fortaleza Cear{\'{a}} Brazil},
        pages        = {1–10},
        _doi          = {10.1145/3132498.3132509},
        _isbn         = {978-1-4503-5325-0},
        _url          = {https://dl.acm.org/_doi/10.1145/3132498.3132509},
        language     = {en}
}

@inproceedings{DBLP:conf/icsa/BuschFK19,
  author       = {Axel Busch and
                  Dominik Fuchss and
                  Anne Koziolek},
  title        = {PerOpteryx: Automated Improvement of Software Architectures},
  booktitle    = {{IEEE} International Conference on Software Architecture Companion,
                , Hamburg, Germany, March 25-26, 2019},
  pages        = {162--165},
  _publisher    = {{IEEE}},
  year         = {2019},
  _url          = {https://_doi.org/10.1109/ICSA-C.2019.00036},
  _doi          = {10.1109/ICSA-C.2019.00036},
  timestamp    = {Thu, 14 Oct 2021 10:38:28 +0200},
  biburl       = {https://dblp.org/rec/conf/icsa/BuschFK19.bib},
  bibsource    = {dblp computer science bibliography, https://dblp.org}
}

@article{Cortellessa-Di-Pompeo-2021,
        title        = {Analyzing the sensitivity of multi-objective software architecture refactoring to configuration characteristics},
        author       = {Cortellessa, Vittorio and {Di Pompeo}, Daniele},
        year         = 2021,
        _month        = {Jul},
        journal      = {Information and Software Technology},
        volume       = 135,
        pages        = 106568,
        _doi          = {10.1016/j.infsof.2021.106568},
        issn         = {09505849},
        language     = {en}
}

@article{DBLP:journals/ec/HildebrandtB15,
  author       = {Torsten Hildebrandt and
                  J{\"{u}}rgen Branke},
  title        = {On Using Surrogates with Genetic Programming},
  journal      = {Evol. Comput.},
  volume       = {23},
  number       = {3},
  pages        = {343--367},
  year         = {2015},
  _url          = {https://_doi.org/10.1162/EVCO\_a\_00133},
  _doi          = {10.1162/EVCO\_A\_00133},
  timestamp    = {Mon, 03 Mar 2025 21:37:25 +0100},
  bib_url       = {https://dblp.org/rec/journals/ec/HildebrandtB15.bib},
  bibsource    = {dblp computer science bibliography, https://dblp.org}
}

@article{franks2008enhanced,
  title={Enhanced modeling and solution of layered queueing networks},
  author={Franks, Greg and Al-Omari, Tariq and Woodside, Murray and Das, Olivia and Derisavi, Salem},
  journal={IEEE Trans. Softw. Eng.},
  volume={35},
  number={2},
  year={2008},
  _publisher={IEEE}
}

@inproceedings{bussemaker2021effectiveness,
  title={Effectiveness of surrogate-based optimization algorithms for system architecture optimization},
  author={Bussemaker, Jasper H and Bartoli, Nathalie and Lefebvre, Thierry and Ciampa, Pier Davide and Nagel, Bj{\"o}rn},
  booktitle={AIAA Aviation 2021 forum},
  pages={3095},
  year={2021}
}

@article{archopt_strategies,
  title={System architecture optimization strategies: dealing with expensive hierarchical problems},
  author={Bussemaker, Jasper H and Saves, Paul and Bartoli, Nathalie and Lefebvre, Thierry and Lafage, R{\'e}mi},
  journal={Journal of Global Optimization},
  volume={91},
  number={4},
  pages={851--895},
  year={2025},
  _publisher={Springer}
}

@inproceedings{archopt_hidden,
  title={Surrogate-based optimization of system architectures subject to hidden constraints},
  author={Bussemaker, Jasper H and Saves, Paul and Bartoli, Nathalie and Lefebvre, Thierry and Nagel, Bj{\"o}rn},
  booktitle={AIAA AVIATION FORUM AND ASCEND 2024},
  pages={4401},
  year={2024}
}

@inproceedings{wang2012baselines,
  title={Baselines and bigrams: Simple, good sentiment and topic classification},
  author={Wang, Sida I and Manning, Christopher D},
  booktitle={Proceedings of the 50th ACL},
  pages={90--94},
  year={2012}
}

@book{jurafsky_speech_2024,
	edition = {Third Edition draft},
	title = {Speech and {Language} {Processing}: {An} {Introduction} to {Natural} {Language} {Processing}, {Computational} {Linguistics}, and {Speech} {Recognition}},
	_url = {https://web.stanford.edu/~jurafsky/slp3/},
	author = {Jurafsky, Daniel and Martin, James H.},
	year = {2024},
}

@article{Leslie2001TheSK,
  title={The Spectrum Kernel: A String Kernel for SVM Protein Classification},
  author={Christina S. Leslie and Eleazar Eskin and William Stafford Noble},
  journal={Pacific Symposium on Biocomputing},
  year={2001},
  pages={
          564-75
        },
  _url={https://api.semanticscholar.org/CorpusID:9725578}
}

@inproceedings{DBLP:conf/qest/LiAZCP17,
	title        = {Tulsa: {A} Tool for Transforming {UML} to Layered Queueing Networks for Performance Analysis of Data Intensive Applications},
	author       = {Chen Li and Taghreed Altamimi and Mana Hassanzadeh Zargari and Giuliano Casale and Dorina C. Petriu},
	year         = 2017,
	booktitle    = {14th {QEST}},
	_publisher    = {Springer},
	series       = {LNCS},
	volume       = 10503,
	pages        = {295--299},
	_doi          = {10.1007/978-3-319-66335-7\_18},
	__url          = {https://_doi.org/10.1007/978-3-319-66335-7\_18},
	_editor       = {Nathalie Bertrand and Luca Bortolussi}
}

@article{DBLP:journals/infsof/CortellessaPST23,
	title        = {Many-objective optimization of non-functional attributes based on refactoring of software models},
	author       = {Vittorio Cortellessa and Daniele {Di Pompeo} and Vincenzo Stoico and Michele Tucci},
	year         = 2023,
	journal      = {Inf. Softw. Technol.},
	volume       = 157,
	pages        = 107159,
	_doi          = {10.1016/j.infsof.2023.107159},
	__url          = {https://_doi.org/10.1016/j.infsof.2023.107159}
}

@inbook{Herold-Klus-Welsch-Deiters-Rausch-Reussner-Krogmann-Koziolek-Mirandola-Hummel,
	title        = {CoCoME - The Common Component Modeling Example},
	author       = {Herold, Sebastian and Klus, Holger and Welsch, Yannick and Deiters, Constanze and Rausch, Andreas and Reussner, Ralf and Krogmann, Klaus and Koziolek, Heiko and Mirandola, Raffaela and Hummel, Benjamin and Meisinger, Michael and Pfaller, Christian},
	year         = 2008,
	booktitle    = {The Common Component Modeling Example},
	_publisher    = {Springer Berlin Heidelberg},
	_address      = {Berlin, Heidelberg},
	series       = {LNCS},
	volume       = 5153,
	pages        = {16–53},
	_doi          = {10.1007/978-3-540-85289-6_3},
	_isbn         = {978-3-540-85288-9},
	issn         = {0302-9743, 1611-3349},
	_url          = {http://link.springer.com/10.1007/978-3-540-85289-6_3},
	editor       = {Rausch, Andreas and Reussner, Ralf and Mirandola, Raffaela and Plášil, František},
	collection   = {LNCS},
	language     = {en}
}

@article{Cao-Smucker-Robinson-2015,                                                                                                                                                                                                            
        title        = {On using the hypervolume indicator to compare Pareto fronts: Applications to multi-criteria optimal experimental design},
        author       = {Cao, Yongtao and Smucker, Byran J. and Robinson, Timothy J.},
        year         = 2015,
        _month        = {May},
        journal      = {Journal of Statistical Planning and Inference},
        volume       = 160,
        pages        = {60–74},
        _doi          = {10.1016/j.jspi.2014.12.004},
        issn         = {03783758},
        language     = {en}
}

@article{Beume-Naujoks-Emmerich-2007,                                                                                                                                                                                                          
        title        = {SMS-EMOA: Multiobjective selection based on dominated hypervolume},
        author       = {Beume, Nicola and Naujoks, Boris and Emmerich, Michael},
        year         = 2007,
        _month        = {Sep},
        journal      = {Eur. J. Oper. Res.},
        volume       = 181,
        number       = 3,
        pages        = {1653–1669},
        _doi          = {10.1016/j.ejor.2006.08.008},
        language     = {en}
}

@inproceedings{Ishibuchi-Masuda-Tanigaki-Nojima-2015,                                                                                                                                                                                          
        title        = {Modified Distance Calculation in Generational Distance and Inverted Generational Distance},
        author       = {Ishibuchi, Hisao and Masuda, Hiroyuki and Tanigaki, Yuki and Nojima, Yusuke},
        year         = 2015,
        booktitle    = {Evolutionary Multi-Criterion Optimization},
        _publisher    = {Springer International Publishing},
        _address      = {Cham},
        pages        = {110–125},
        _doi          = {10.1007/978-3-319-15892-1_8},
        _url          = {http://link.springer.com/10.1007/978-3-319-15892-1_8},
        _editor       = {Gaspar-Cunha, António and Henggeler Antunes, Carlos and Coello, Carlos Coello}
}

@article{Li-Yao-2020,                                                                                                                                                                                                                          
        title        = {Quality Evaluation of Solution Sets in Multiobjective Optimisation: A Survey},
        author       = {Li, Miqing and Yao, Xin},
        year         = 2020,
        _month        = {Mar},
        journal      = {ACM Comput. Surv.},
        volume       = 52,
        number       = 2,
        pages        = {1–38},
        _doi          = {10.1145/3300148},
        issn         = {0360-0300, 1557-7341},
        language     = {en}
}

@inproceedings{DBLP:conf/icsa/CortellessaP024,
  author       = {Vittorio Cortellessa and
                  Daniele {Di Pompeo} and
                  Michele Tucci},
  title        = {Exploring Sustainable Alternatives for the Deployment of Microservices
                  Architectures in the Cloud},
  booktitle    = {21st {IEEE} International Conference on Software Architecture, {ICSA}},
  pages        = {34--45},
  _publisher    = {{IEEE}},
  year         = {2024},
  _url          = {https://_doi.org/10.1109/ICSA59870.2024.00012},
  _doi          = {10.1109/ICSA59870.2024.00012},
  timestamp    = {Tue, 30 Jul 2024 14:08:48 +0200},
  biburl       = {https://dblp.org/rec/conf/icsa/CortellessaP024.bib},
  bibsource    = {dblp computer science bibliography, https://dblp.org}
}

@article{DBLP:journals/software/Zimmermann15,
  author       = {Olaf Zimmermann},
  title        = {Architectural Refactoring: {A} Task-Centric View on Software Evolution},
  journal      = {{IEEE} Softw.},
  volume       = {32},
  number       = {2},
  pages        = {26--29},
  year         = {2015},
  _url          = {https://_doi.org/10.1109/MS.2015.37},
  _doi          = {10.1109/MS.2015.37},
  timestamp    = {Mon, 08 Jun 2020 22:31:27 +0200},
  biburl       = {https://dblp.org/rec/journals/software/Zimmermann15.bib},
  bibsource    = {dblp computer science bibliography, https://dblp.org}
}

@article{Arcuri-Fraser-2013,
        title        = {Parameter tuning or default values? An empirical investigation in search-based software engineering},
        author       = {Arcuri, Andrea and Fraser, Gordon},
        year         = 2013,
        _month        = {Jun},
        journal      = {Empirical Software Engineering},
        volume       = 18,
        number       = 3,
        pages        = {594–623},
        _doi          = {10.1007/s10664-013-9249-9},
        issn         = {1382-3256, 1573-7616},
        language     = {en}
}

@article{DBLP:journals/tec/DebJ14,
  author       = {Kalyanmoy Deb and
                  Himanshu Jain},
  title        = {An Evolutionary Many-Objective Optimization Algorithm Using Reference-Point-Based
                  Nondominated Sorting Approach, Part {I:} Solving Problems With Box
                  Constraints},
  journal      = {{IEEE} Trans. Evol. Comput.},
  volume       = {18},
  number       = {4},
  pages        = {577--601},
  year         = {2014},
  _url          = {https://_doi.org/10.1109/TEVC.2013.2281535},
  _doi          = {10.1109/TEVC.2013.2281535},
  timestamp    = {Tue, 12 May 2020 16:50:49 +0200},
  biburl       = {https://dblp.org/rec/journals/tec/DebJ14.bib},
  bibsource    = {dblp computer science bibliography, https://dblp.org}
}

\end{document}